\documentclass[preprint,rmp,amsmath,amssymb,showkeys,floatfix]{revtex4}
\usepackage{graphicx}

\newcommand{\cut}[1]{}

\newcommand{\bleq}{\ifpreprintsty \else
\end{multicols}\vspace*{-3.5ex}{\tiny \noindent\begin{tabular}[t]{c|}
\parbox{0.493\hsize}{~} \\ \hline \end{tabular}} \fi}
\newcommand{\eleq}{\ifpreprintsty \else
{\tiny\hspace*{\fill}\begin{tabular}[t]{|c}\hline
\parbox{0.49\hsize}{~} \\
\end{tabular}}\vspace*{-2.5ex}\begin{multicols}{2} \fi}

\def\Nkhooke{k_{\rm sp}}% hooke's law const
\def\Neffsp{\kappa}     %effective spring const
\def\Nxterm{\eta}          %crossterm
\def\Nestiff{E}         %extensional stiffness
\def\Nlaest{y}          %estimated e-value
\def\kb{{k_{\rm B}}}
\def\kbt{{k_{\rm B}T}}
\def\fjc{{\ensuremath{\mathrm{FJC}}}}
\def\wlc{{\ensuremath{\mathrm{WLC}}}}
\def\dpc{{\ensuremath{\mathrm{DPC}}}}
\def\tot{_{\mathrm{tot}}}
\def\eref#1{Eq.~\ref{#1}}
\def\erefs#1{Eqs.~#1} %if you have several
\def\sref#1{Sect.~\ref{#1}}
\def\fref#1{Fig.~\ref{#1}}
\def\pNunit{\ensuremath{\mathrm{pN}}}
\def\nmunit{\ensuremath{\mathrm{nm}}}
\def\umunit{\ensuremath{\mu\mathrm{m}}}
\def\that{\hat t}
\def\Tmat{{\sf T}}
\newcommand{\inv}{^{\raise.15ex\hbox{${\scriptscriptstyle -}$}\kern-.05em 1}}

\newcommand{\normsq}[1]{\|#1\|^2}

\begin{document}

\title{Theory of High-Force DNA Stretching and Overstretching}

\author{Cornelis Storm}
\author{Philip Nelson}
\affiliation{Department of Physics and Astronomy, University of Pennsylvania,
Philadelphia, Pennsylvania 19104 USA}

\date{\today}
\begin{abstract}
Single molecule experiments on single- and double stranded DNA
have sparked a renewed interest in the force-extension of
polymers. The extensible Freely Jointed Chain (FJC) model is  frequently
invoked to explain the
observed behavior of single-stranded DNA. We demonstrate that this model
does not satisfactorily describe recent
high-force stretching data. We instead propose a model (the \textit{Discrete
Persistent Chain,} or ``DPC'') that borrows features from both the FJC
and the Wormlike Chain, and show that it resembles the data more
closely. We find that most of the high-force behavior previously
attributed to stretch elasticity is really a feature of the corrected
entropic elasticity; the true stretch compliance of single-stranded DNA is
several times smaller than that found by previous authors. Next we
elaborate our model to allow coexistence of two
conformational states of DNA, each with its own stretch and bend elastic
constants.
Our model is computationally simple, and gives an excellent fit through the
entire
overstretching transition of nicked, double-stranded DNA. The fit  gives
the first values for the elastic constants of the stretched state.
In particular we find the effective bend stiffness for DNA in this state
to be about $10\,\mathrm{nm}\cdot\kbt$, a value quite different from
either B-form or
single-stranded DNA.
\end{abstract}
\keywords{Nucleic acid conformations}
\maketitle

\section{Introduction and Summary\label{s:IS}}
New single-molecule manipulation techniques have opened the
mechanical properties of individual macromolecules to much more direct
study than ever before. For example, optical-trap measurements give
the force-extension relation of a single molecule of lambda DNA, from which we
can deduce the molecule's average elastic properties by fitting to a model.
Part of the beauty of this procedure is that we pass from an
optical-scale measurement (the total end-to-end length of the DNA is
typically over 10$\,\umunit$) to a microscopic conclusion (the elastic
constants of the $2\,\nmunit$ diameter DNA molecule). But by the same
token, we must be careful with the interpretation of our results.
Fitting a physically inappropriate model to data can give
reasonable-looking fits, but yield values of the fit
parameters that are not microscopically meaningful.

We will illustrate the above remarks by studying  high-force
measurements of the force-extension relation for single-stranded DNA.
Previous authors have fit this relation at  low to moderate forces to
the Extensible Freely Jointed Chain (EFJC) model, obtaining as fit parameters
a link length and an extension modulus for increasing the contour length
of the chain. We argue that to capture the microscopic physics, at
least one additional element must be added to the model, namely a link
stiffness. The resulting model fits the data better than either the
EFJC or the Extensible Worm-Like Chain (EWLC) models. The fit also yields a
much large value of the extension modulus than previously reported. The
reason for this discrepancy is that high-force effects previously
attributed to intrinsic stretching of the chain are, in our model,
simply a part of the correct entropic elasticity.

The mathematical formalism we introduce to solve our model is of some
independent interest, being simpler than some earlier approaches.
In particular, it is quite easy to extend our model to study a linear
chain consisting of two different, coexisting conformations of the
polymer, each with its own elastic constants.  We formulate and solve
this model as well.  The model makes no assumptions about the elastic
properties of the two states, but rather deduces them by fitting to
recent data on the overstretching transition in nicked,
double-stranded DNA. Besides giving a very good fit to the data, our
model yields insight into the character of the stretched conformation
of DNA. The model is flexible and can readily be adapted to the study
of the stretching of polypeptides with a helix-coil transition.

\cut{The advent of experimental devices such as the atomic force microscope
(AFM) and the optical tweezers (OT) has greatly improved our
understanding of the behavior of polymers.  They allow for precise
studies at the level of single polymers, providing a near-ideal
testing ground for a variety of theories circulating in the polymer
community.  As such, they have contributed to new insights into the
elasticity and rheology of polymers, their solutions and their
networks.  The AFM and OT setups are particularly well-suited to exert
tiny (in the picoNewton range) but well-controlled forces on single
macromolecules and measure their response.  Knowledge of the
force-extension curves can help gain insight into the structure of
folded proteins or the elasticity of polymers, entropic and beyond.}
\section{The Worm-Like Chain and the Freely Jointed Chain}
\subsection{The Freely Jointed Chain}
A polymer is a long, linear, single molecule. The chemical bonds
defining the molecule can be more or less flexible in different cases.
The simplest model of polymer conformation treats the molecule as a
chain of rigid subunits, joined by perfectly flexible hinges---a ``freely
jointed chain,'' or FJC \cite{Flory}. The FJC model is not very appropriate
to double-stranded DNA, consisting of a stack of flat basepairs joined
by both covalent bonds and  physical interactions (hydrogen bonds and
the hydrophobic base-stacking energy), but for single-stranded DNA (ssDNA) it
forms an attractive starting point.

Deviations from the FJC picture can come from a variety of interactions
among the individual monomers: Individual covalent bonds may have bending
energies that are not small relative to $\kbt$, successive monomers may
have steric interactions, and so on. To some extent we can compensate
for the model's omission of such interactions by choosing an effective
link length $b$ that is longer than the actual monomer size. Since the
FJC views the polymer as a chain of perfectly stiff links, choosing a
larger $b$ gives us a chain of  longer links and thus effectively stiffens the
chain. Accordingly, one views $b$ as a fit parameter when
deriving the force-extension relation of the model. The fit
value of $b$ can then depend both on the molecule under study, and on
its external conditions like salt concentration, as those conditions
affect the intramolecular interactions.

\begin{figure}
\begin{center}
  \includegraphics[width=.8\linewidth]{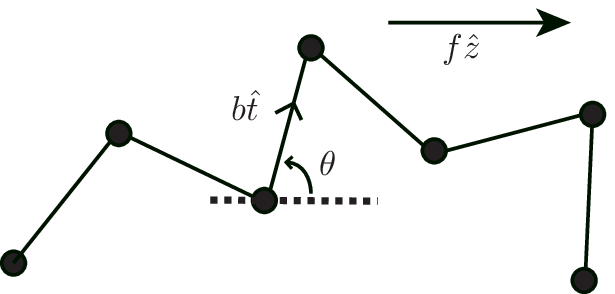}
\end{center}
\caption[]{The Freely Jointed Chain consists of identical segments
of length $b$, joined together by free hinges. The
configuration is fully described by the collection of orientation
vectors $\{\hat t_i\}$.  $\{\theta_i\}$ denotes the angle between
$\that_i$ and the fixed direction $\hat z$ of the applied stretching
force.}\label{fjcfig}
\end{figure}

To formulate the FJC we describe a molecular conformation by
associating with each segment a unit orientation vector $\hat t_i$,
pointing in the direction of the $i$th segment, as sketched in
\fref{fjcfig}.  In the presence of an
external force $\vec{f}$ along the $\hat z$ direction, we can define
an energy functional for the chain
\begin{equation}
\label{ener}
\frac{{\cal E}^{\fjc}[\hat t_i]}{\kb  T}=-\sum_{i=1}^{N}\frac{fb}{\kb  T}
\, \hat t_i \cdot \hat z\, .
\end{equation}
In the absence of an external force, all configurations have equal
energy and (neglecting self-avoidance) the chain displays the
characteristics of a random walk.  To pull the ends of such a chain
away from each other a force has to be applied, as extending the chain
reduces its conformational entropy.  The resulting entropic elastic
behavior can be summarized in the {\em force-extension relation}
\cite{Grosberg}
\begin{equation}
\label{ferfjc}
\langle \frac{z}{L_{\text{tot}}}\rangle = \coth(\frac{fb}{\kb
T})-\frac{\kb  T}{fb} \, ,
\end{equation}
the well-known Langevin function.  In the limit of low stretching
force, all polymer models reduce to Hooke-law behavior $f=\Nkhooke\langle
z\rangle$; we define the
effective spring constant by $\Neffsp=\Nkhooke \cdot L\tot$, or
\begin{equation}
\label{lofo}
\langle\frac{z}{L_{\mathrm{tot}}}\rangle \to \frac{f}{\Neffsp} + {\cal
O}(f^2) \, .
\end{equation}
Expanding \eref{ferfjc} gives the effective spring constant for
the FJC as $\Neffsp^{\text{FJC}}=\tfrac{3 \kbt}{b}$.  The fact that
the effective spring constant is proportional to the absolute
temperature illustrates that the elasticity in this model is purely
entropic in nature.

At high stretching force, \eref{ferfjc} gives $\langle\frac
z{L\tot}\rangle\to1$;
the extension saturates when all the links of the chain are aligned by
the external force. In reality, individual links are slightly
extensible; we will modify  the model to introduce this effect in
\sref{e1}.

\subsection{The Wormlike Chain\label{wlc}}
As mentioned above, double-stranded DNA (dsDNA) is far from being a freely
jointed chain. Thus it is unsurprising that while the FJC model can
reproduce the observed linear force-extension relation of dsDNA at low
stretching force, and the observed saturation at high force, still it fails at
intermediate values of $f$. Another indication that the model is
physically inappropriate is that the best-fit value of the link length
is $b\approx100\,$nm, completely different from the physical contour
length per basepair of $0.34\,$nm.

\begin{figure}
\begin{center}
  \includegraphics[width=.8\linewidth]{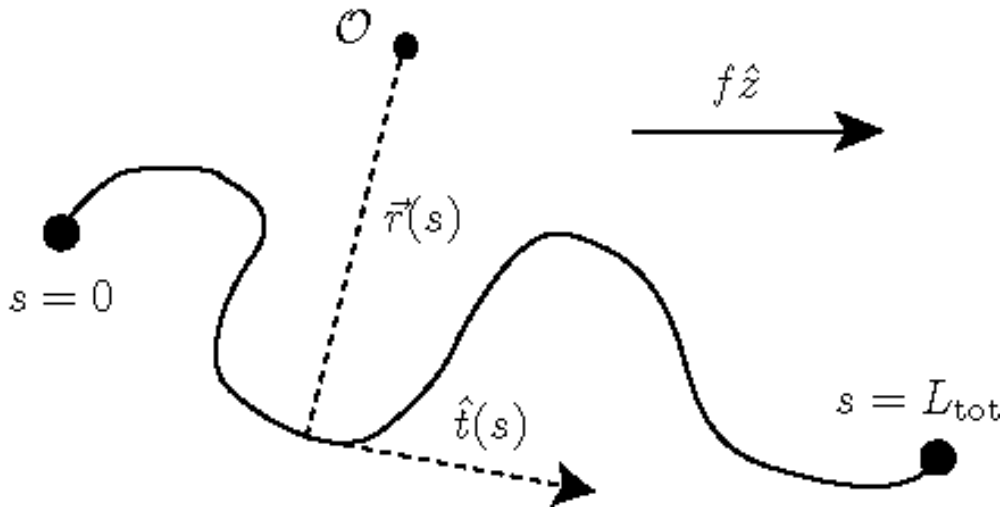}
\end{center}
\caption[]{A Worm Like Chain is a continuum elastic medium, whose
configuration is described in terms of the position vector $\vec
r$ as a function of the contour length $s$.}\label{wlcfig}
\end{figure}

To improve upon the FJC, we must account for the fact that the
monomers {\em do} resist bending. In fact, the very great stiffness of
double-stranded DNA can be  turned to our advantage, as it implies
that successive monomers are constrained to point in nearly the same
direction. Thus we can treat the polymer as a continuum
elastic body, its configuration described by the position $\vec
r(s)$ as a function of the relaxed-state contour length $s$ (see \fref{wlcfig}).
Continuing to treat the chain as inextensible gives the Worm Like
Chain \cite{KratkyPorod,sait67}. The local tangent and curvature
vectors ($\vec t$ and $\vec w$, respectively) are given by
\begin{equation}
\vec t(s)=\frac{{\rm d}\vec r(s)}{{\rm d}s}\, , \quad \vec w(s)=\frac{{\rm
d}\vec t(s)}{{\rm d}s}\, .
\end{equation}
We temporarily assume that the chain is inextensible, expressed locally by the
condition that $|\vec t(s)|=1$ everywhere.

To get an energy functional generalizing \eref{ener}, we note that for
a thin, homogeneous rod the energy density of elastic strain is proportional
to the square of the local curvature. Adding the external-force term
from \eref{ener} yields
\begin{equation}\label{wlcener}
\frac{{\cal E}^{\wlc}[\hat t(s)]}{\kb  T}=\int_0^{L_{\text{tot}}}{\rm
d}s \left\{ \frac{A}{2}\left| \frac{{\rm d}\hat t(s)}{{\rm d}s}\right|^2-\frac{f}{\kb
T}\hat t(s) \cdot \hat z\right\}\, .
\end{equation}
\eref{wlcener} makes it clear that
the parameter
$A$ is a measure of the {bend stiffness} of the chain. $A$ is also the {\em
persistence length} of the chain, the characteristic length scale
associated with the decay of tangent-tangent correlations at zero
stretching force:
\begin{equation}
\label{ttdecay}
\langle \hat t(0) \cdot \hat t(s)\rangle_{\text{WLC}} \sim e^{-|s|/A}\, .
\end{equation}

The force-extension relation for the WLC was obtained numerically in
\cite{MarkoSiggia}; subsequently a high-precision
interpolation formula was given in \cite{Bouchiat}.  At low force, the WLC also behaves like an ideal
spring, with effective spring constant \cite{Yamakawabook}
\begin{equation}
\label{kspwlc}
\Neffsp^{\text{WLC}}=\frac{3 \kb  T}{2 A}\, .
\end{equation}
Thus a WLC with stiffness parameter $A$ yields a force-extension
relation that at low force matches the FJC with $b=2A$.

The remarks at the start of this subsection make it clear that
the WLC is just an approximation, valid in the limit where the
persistence length $A$ is much longer than the physical monomer length
(and width). When these conditions are not met, the picture of the
molecule as a thin, continuous, elastic body will not be accurate;
short-length {cutoff} effects will then enter in an essential way.

\subsection{Experiments\label{e1}}
Early single-molecule stretching experiments showed that
double-stranded DNA closely follows the predicted force-extension of
the WLC at forces under $10\,$pN \cite{bust94a}. Later experiments
probing the region $10\,\pNunit<f<60\,\pNunit$ found a linear
deviation from the WLC prediction, attributable to a Hooke-law
stretching elasticity \cite{Cluzel,Smith,Awang97a}. Adding this
effect into the model introduces a second fit parameter $\Nestiff$ in
addition to $A$. To lowest order in $f/\Nestiff$ this modification just
amounts to multiplying the model's
$\langle\frac z{L\tot}\rangle$ by the factor
$(1+\tfrac{f}{\Nestiff })$; for dsDNA the resulting fit is very good out to
$60\,\pNunit$.

\begin{figure}
\begin{center}
  \includegraphics[width=.8\linewidth]{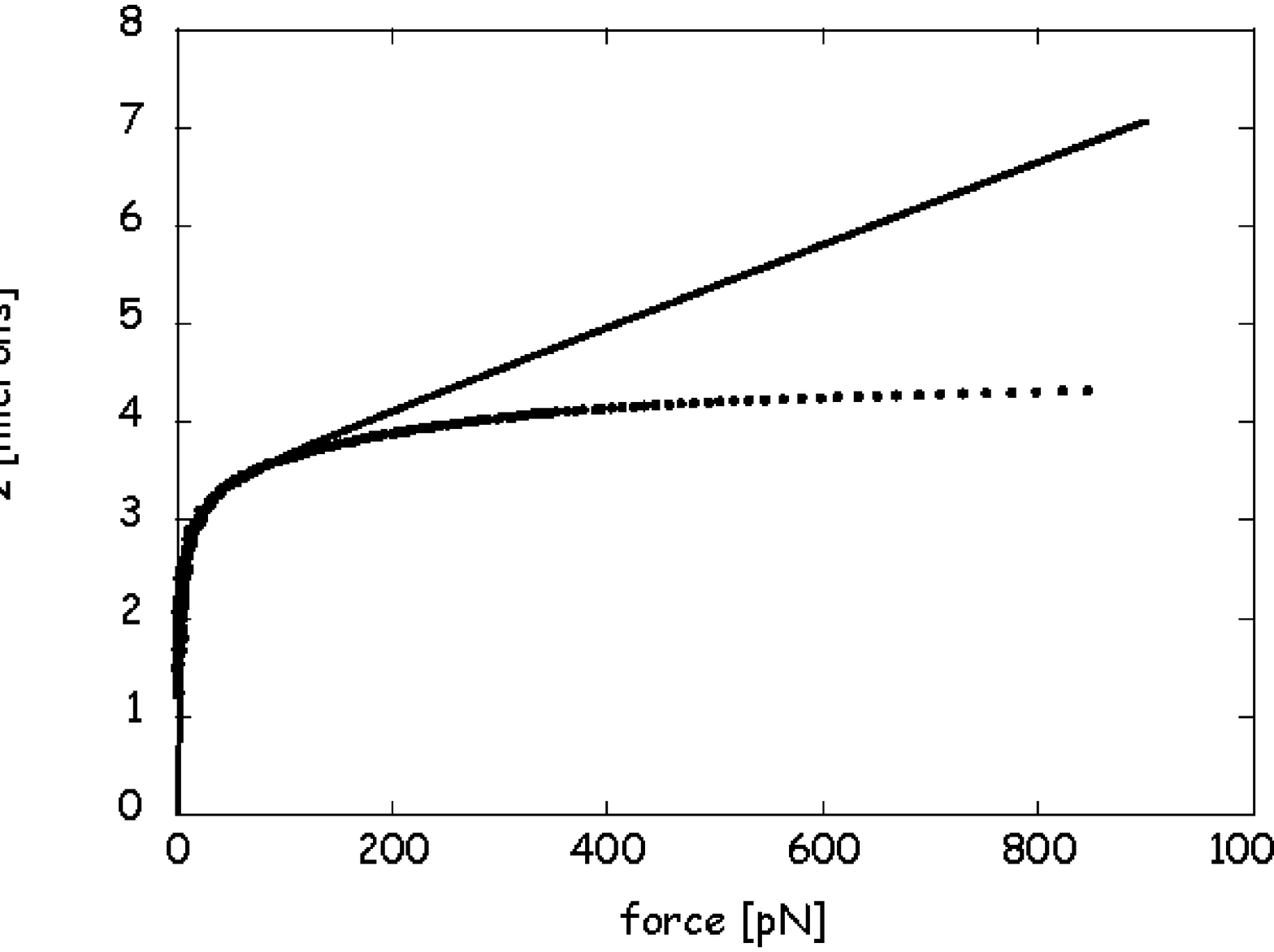}
\end{center}
\caption[]{Least-squares fit (solid line) of the
single-stranded DNA stretching data (closed circles) from \cite{Rief} to
the extensible FJC model.  Included in the fit are
the data up to a force of 100\thinspace pN. Fitting only those data points
yields
a link length $b=1.75\,$nm and a stretch modulus $\Nestiff =8\cdot
10^2\,\pNunit$, reproducing the typical values as cited for instance in
\cite{Rief,Clausen,Hegner}.  In this graph, we have extrapolated this
fit to the high-force range, to demonstrate that the parameters as
extracted from the low-force data do not represent the full range of
data faithfully.}\label{fjcextra}
\end{figure}

The situation for single-stranded DNA has been less clear.  Adding an
extensibility factor to \eref{ferfjc} again yields a model with two
parameters ($b$ and $\Nestiff$).  Though this ``extensible FJC'' (EFJC) model
yielded impressive fits to the early experimental data, recent
advances in single-molecule manipulation \cite{Rief,Clausen} have
again probed higher forces, and here the agreement is not so good.  As
shown in \fref{fjcextra}, the previously cited values for $b$ and
$\Nestiff $ do not give a successful extrapolation  to the regime of higher
forces.  In the following section, we will propose a new model that
borrows features from both the FJC and the WLC to describe these
data more accurately.

\section{The Discrete Persistent Chain\label{secdpc}}
The previous sections have made it clear that a real polymer will
display \textit{both} discreteness and bend stiffness. While
we have seen that the corresponding effects on the force-extension
relation are interchangeable at very low forces, higher forces {\em will}
distinguish them. Accordingly we now formulate a model with {\em both} $b$ and
$A$; later we will add a stretch stiffness as well.

\begin{figure}
\begin{center}
  \includegraphics[width=.8\linewidth]{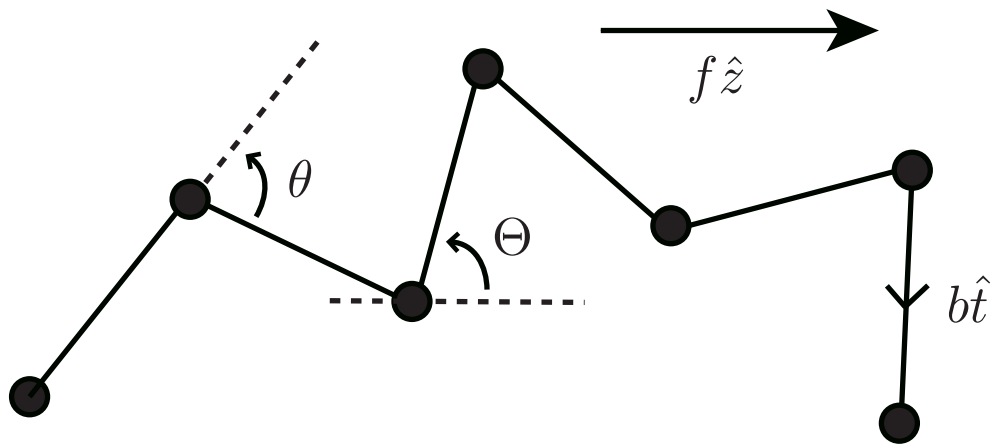}
\end{center}
\caption[]{The Discrete Persistent Chain, viewed as a FJC with
an additional term in the energy proportional to the square of the
polar angle $\Theta$ between successive segments.}\label{dpcfig}
\end{figure}

Our ``Discrete Persistent Chain'' (or DPC) models the polymer as a chain
composed of $N$ segments
of length $b$, whose conformation is once again fully described by
the collection of orientation vectors $\{\hat t_i\}$ (see \fref{dpcfig}). Bend
resistance is taken into account by including an energy penalty at
each link proportional to the square of the angle
($\Theta_{i,i+1}=\arccos\hat t_i \cdot \hat t_{i+1}$) between two subsequent
links. The energy functional describing this model is thus given by
\begin{equation}
\frac{{\cal E}^{\dpc}[\{\hat t_i\}]}{\kb  T}=-\sum_{i=1}^{N}\frac{fb}{\kb
                     T} \, \hat t_i \cdot \hat z\,
+\sum_{i=1}^{N-1}\frac{A}{2b}(\Theta_{i,i+1})^2  \, .
\end{equation}
The partition function for this energy functional is then given by
\begin{equation}\label{pfdpc}
{\cal Z}=\left[ \prod_{i=1}^{N} \int_{{\mathbb S}^2} \! {\rm d}^2\hat t_i
\!\right] e^{-\frac{fb}{2\kb T} \hat t_1 \cdot \hat z}\left\{\prod_{i=1}^{N-1}
e^{-{\cal E}_i(\hat t_i,\hat t_{i+1})/\kb  T}\right\}e^{-\frac{fb}{2\kb T} \hat t_N
\cdot \hat z}
\, ,
\end{equation}
where
\begin{equation}\label{pfdpcb}
\frac{{\cal E}_i(\hat t_i,\hat t_{i+1})}{\kb T}=-\frac{fb}{2 \kb T} \, (\hat
t_i+\hat t_{i+1}) \cdot \hat z\, +\frac{A}{2b}(\Theta_{i,i+1})^2
\end{equation}
and ${\mathbb S}^2$ is the two-dimensional unit sphere.

To compute ${\cal Z}$ we interpret each integral in \eref{pfdpc} as a
generalized matrix product (among matrices with continuous indices),
writing \cite{KramersWannier}
\begin{equation}
{\cal Z}=\vec v \cdot {\sf T}^{N-1}\cdot \vec w \, .
\end{equation}
In this formula $\vec v$ and $\vec w$ are vectors indexed by $\hat t$,
or in other words functions $v(\that\,),w(\that\,)$. The matrix product
$\Tmat\cdot\vec v$ is a new vector, defined by the convolution:
\begin{equation}
\label{dp1}
({\sf T}\cdot \vec v\,)(\hat t_i)=\int_{{\mathbb S}^2} \! {\rm d}^2\hat t_j
\, {\mathbb T}(\hat t_i,\hat t_j) v(\hat t_j) \, .
\end{equation}
The matrix elements of $\Tmat$ are given by
\begin{equation}\label{opent}
{\mathbb T}(\hat t_i,\hat t_j)=e^{-{\cal E}_i(\hat t_i,\hat t_j)/\kb  T}\, ;
\end{equation}
we will not need the explicit forms of $\vec v$ and $\vec w$ below.

The force-extension relation can be obtained from ${\cal Z}$ by
differentiating with respect to the force (see
\erefs{\ref{pfdpc}--\ref{pfdpcb}}):
\begin{equation}
\langle \frac{z}{L_{\text{tot}}} \rangle=\left( \frac{\kb
T}{L_{\text{tot}}} \right) \frac{\rm d}{{\rm d} f} \ln {\cal Z}\,.
\end{equation}
It is here that the transfer matrix formulation can be used to greatly
simplify the calculation of the force-extension relation, since all
that is needed to compute the logarithmic derivative of ${\cal Z}$ in
the limit of long chains is the largest eigenvalue of ${\sf T}$, which
we will call $\lambda_{\text{max}}$:
\begin{equation}
\label{zldef}
\langle \frac{z}{L_{\text{tot}}} \rangle\stackrel{\mathrm{large
}N}{\longrightarrow}\left(\frac{\kb  T}{L_{\text{tot}}}\right) \frac{\rm
d}{{\rm d} f}
\ln (\lambda_{\max})^{N}
=\left(\frac{\kb  T}{b}\right) \frac{\rm d}{{\rm d} f} \ln {\lambda_{\max}}\, .
\end{equation}

\begin{figure}
\begin{center}
  \includegraphics[width=.8\linewidth]{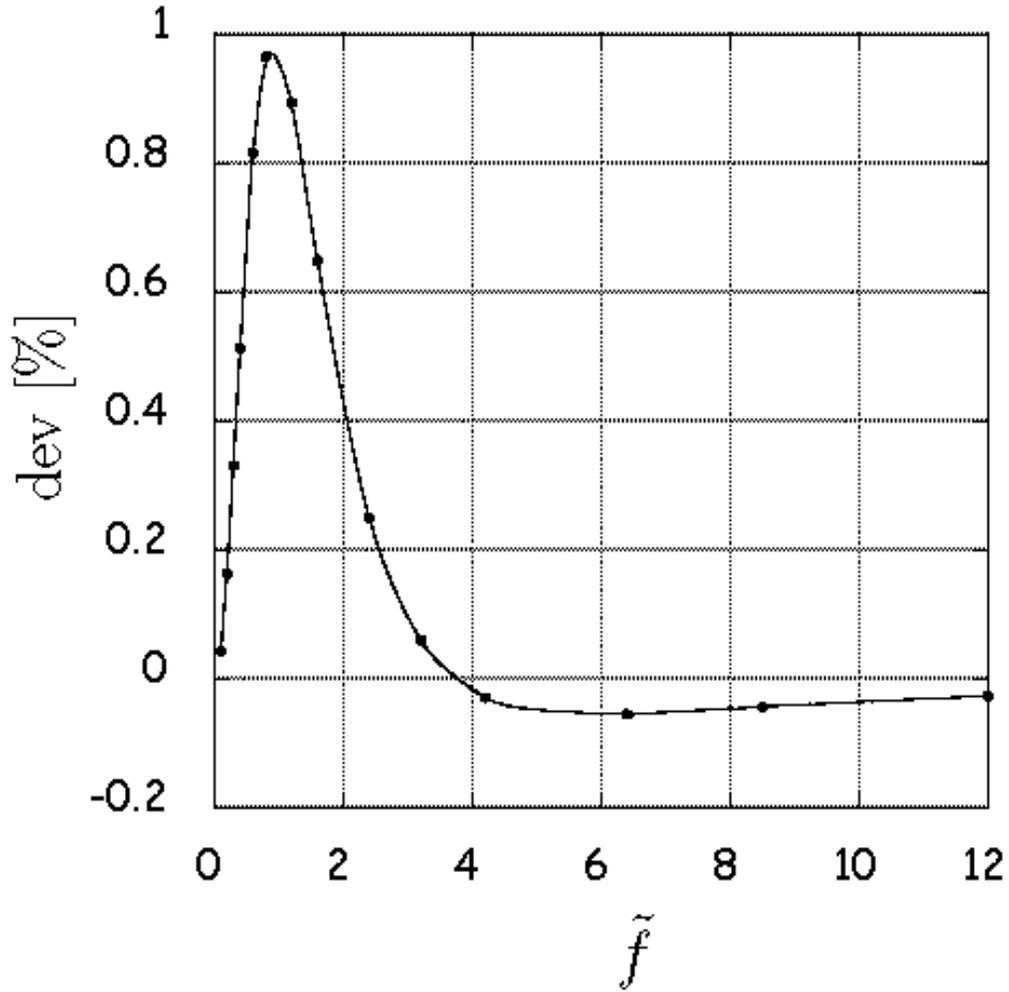}
\end{center}
\caption[]{Comparison between the exact WLC force-extension solution and
the Ritz variational approximation. The deviation ${\rm dev}(\tilde f)$ is
defined as $100\% \times(z(\tilde f)_{\rm exact}\!-\!z(\tilde f)_{\rm
var})/z(\tilde f)_{\rm exact}$, with $\tilde f$ the dimensionless force
$\tilde f= \frac{f A}{\kb  T}$. The maximal error induced by the
variational approximation is about $1\%$. Data for the exact solution were
taken from \cite{Bouchiat}.}\label{compare}
\end{figure}

We will approximate $\lambda_{\text{max}}$ using a variational
scheme. Following the line of argument of \cite{MarkoSiggia}, we note that
the leading eigenfunction of ${\sf
T}$ will reflect the physics of the problem in the sense that it must
be azimuthally symmetric and peaked in the direction of
the applied force.  A suitable 1-parameter family of trial
eigenfunctions $\vec v_{\omega}$ can therefore be defined by
\begin{equation}
 v_\omega (\hat t\,)=e^{\omega \hat t \cdot \hat z}\, .
\end{equation}
Under (\ref{dp1}), the  $\vec v_{\omega}$ have squared norms
\begin{equation}
\label{vnorm}
\|\vec v_{\omega}\|^2=\tfrac{2 \pi}{\omega} \sinh(2 \omega)\, ,
\end{equation}
which allows us to approximate $\lambda_{\text{max}}$ variationally by
\begin{equation}
\label{laestdef}
\lambda_{\text{max}}^*\equiv\max_\omega \Nlaest (\omega)\equiv\max_\omega \,
\frac{\vec v_{\omega} \cdot {\sf T} \cdot \vec v_{\omega}}{\normsq{\vec
v_\omega}}
\,.\end{equation}
To get some idea of the quality of this variational approach, we can
compare its results in the limit $b\to0$ (the WLC) to the exact
solution of that model.  \fref{compare} plots the difference of
these
force-extension curves, and shows that the results from the
variational approximation are nowhere off by more than $1\%$.

Returning to the full DPC model, Appendix~\ref{appx} shows that it is
possible to express $\Nlaest(\omega)$ in terms of the
dimensionless variables
\begin{equation}
\tilde f=\frac{f b}{\kb  T}\,, \qquad \tilde \ell = \frac{A}{b}
\end{equation}
as a combination of error functions as follows
\begin{eqnarray}\label{laest1}
\Nlaest(\omega)& = & \frac{2 \sqrt 2 \pi^{3/2} \omega e^{-2 \tilde \ell
-\frac{(2\omega+\tilde f)^2}{8\tilde \ell}}\text{csch}(2
\omega)}{\sqrt{-\tilde \ell}(2\omega+\tilde f)} \times \nonumber \\
 & &\hspace{-.5cm} \times \left[ \text{Erf}\left( \frac{i}{2\sqrt{2 \tilde
\ell}} (\tilde f + 4 \tilde \ell + 2 \omega)\right)-\text{Erf}\left(
\frac{i}{2\sqrt{2 \tilde \ell}} (\tilde f - 4 \tilde \ell + 2 \omega)
\right)\right] \, .
\end{eqnarray}
This formula is only valid in the parameter regime where $\omega^*$ (the
locus of
the maximum of \eref{laest1}) obeys
\begin{equation}\label{valid}
\omega^*>\tilde \ell-\frac12 \tilde f\, .
\end{equation}
For practical purposes this is the region where the magnitude of the
bend stiffness $A$ is larger than, or at most comparable to, the link
length $b$, which is the physically relevant regime. We maximize
\eref{laest1} numerically to obtain
$\lambda_{\text{max}}^*$, from which we can then compute the
force-extension relation by numerical differentiation with respect to
the force.  In the small force limit, we can do a little better based
on the observation that for small $\tilde f$, $\omega^*$ is also small.
Expanding
\eref{laest1} to second order in $\omega$ and $\tilde f$ we can
analytically solve the stationarity condition $\frac{{\rm
d}\Nlaest}{{\rm d} \omega}=0$ (which is now simply a
quadratic equation) and determine the small force entropic elastic
behavior of our DPC model to be
\begin{equation}
\langle \frac{z}{L_{\text{tot}}}\rangle\to\frac f{\Neffsp^{\dpc}}+{\cal
O}(f^2)\cut{
 \frac{2 A^{\text{DPC}}_{\text{eff}}}{3 \kb  T}f}\, ,
\end{equation}
where the {effective} spring constant for the DPC model is given
by\footnote{\eref{screnorm} has the expected property that
$\Neffsp^{\dpc}\to\Neffsp^{\wlc}$ when we send $b\to 0$ with $A$ fixed. The
opposite limit, where $A$ goes to $0$ holding $b$ fixed, should recover the
FJC, but instead \eref{screnorm} gives an unphysical, {\em negative} value
of $\Neffsp^{\dpc}$. However, this limit takes us outside the domain of
validity \eref{valid}, and we cannot use \eref{screnorm} any more. We have
verified numerically that the DPC model {\em does} reduce to the FJC in
that particular limit.}
\begin{equation}\label{screnorm}
\Neffsp^{\dpc}=\frac32\frac{\kbt}A\left(1-\frac b{2A}\right)\inv
\cut{ A^{\text{eff}}=A(1-\frac{b}{2 A}) }\, .
\end{equation}
It is sometimes convenient to reexpress the parameters $A$ and $b$ of
the DPC model in terms of $\Neffsp^{\dpc}$ and $b$. We do this using \eref{screnorm}:
\begin{equation}
A=\frac{b}{2}+\frac{3 \kb T}{2 \Neffsp^{\dpc}}\, .
\end{equation}

\begin{figure}
\begin{center}
  \includegraphics[width=.8\linewidth]{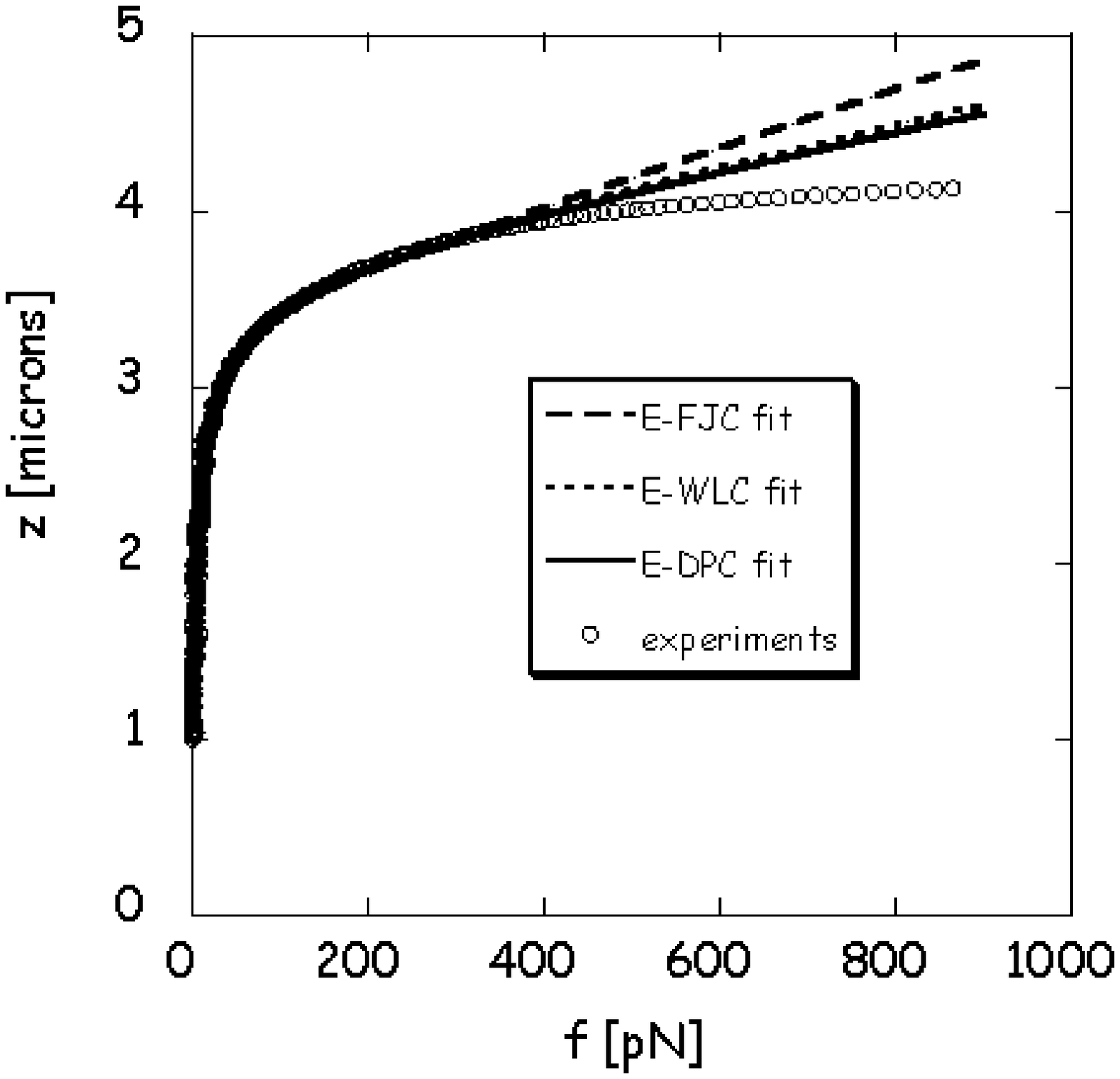}
\end{center}
\caption[]{Fit of the extensible DPC model (solid line) to
the single-strand DNA stretching data (circles) kindly supplied by M. Rief;
see \cite{Rief}.  The fit shown was obtained for $b=0.17\, ,\Nestiff
=4.5\cdot 10^3\,\pNunit\, , L_{\rm tot}=3.9\, \mu$m, and
$\Neffsp^{\dpc}=\frac32\frac{\kbt}{0.85\,\mathrm{nm}}$. In addition, the
dashed and dotted lines show the corresponding best fits to the
extensible FJC and WLC, respectively. All fits include the data points
only for
forces between 20 pN and 400 pN. Values for $\chi^2$ were EFJC :
$\chi^2=1.269$; EWLC : $\chi^2=0.600$ and EDPC : $\chi^2=0.490$ at
$N=1523$. We ignore the lowest-force points because
of complications induced by hairpins and other secondary
structures in the DNA. }\label{compareall}
\end{figure}

It is straightforward to add an intrinsic stretch modulus to the
calculation outlined above, obtaining the ``Extensible DPC'' (or EDPC)
model.
We have computed the resulting force-extension curves, and fitted to recent
data for ssDNA. The results of
these fits are collected in \fref{compareall}.  Fitting to the data points
with $f<400\,$pN yields
a value of the stretch modulus of around
$\Nestiff\approx4500\,\pNunit$, more than four times larger than even
the largest of the previous estimates \cite{Rief,Clausen,Hegner}. We
interpret this discrepancy by noting that if we hold $\Neffsp$ constant
while varying $b$, the difference between the EFJC and EDPC models
shows up in the high-force regime, which is also sensitive to the
choice of $\Nestiff$.  Thus neglecting cutoff effects causes curve
fitting to make a compensating change in $\Nestiff$.

The best fit (in terms of $\chi^2$) is obtained for a
value of $b\approx0.17\,$nm, away from both the EWLC ($b=0$) and EFJC
($b=\tfrac{3 \kb T}{\Neffsp}=1.7\,$nm) limits of the
model. Even though to the eye the difference between the three models in
the fit region might appear marginal, the improvement in $\chi^2$ achieved
by the DPC at just over $18\%$ is statistically relevant. Interestingly,
the fit value of $b$ is indeed comparable
to the physical segment length of ssDNA ($0.6\,$nm), a result not put
in by hand. \fref{compareall} also shows that our EDPC model extrapolates
better
to the high-force regime than either the EFJC or the EWLC.

Previous authors have already noted that the extensible FJC model does not
accurately model the high-force data \cite{Rief,Clausen}, but have
attributed its failure
to the onset of nonlinear elasticity effects. We may expect such
effects to become significant when the ratio $f/\Nestiff$ exceeds, say,
10\%. Our large fit value of $\Nestiff$ means that we ought to be able
to trust our linear-elasticity model out to around $f=400\,\pNunit$,
which is why we used only the data up to this point in our fit.
Indeed \fref{compareall} shows that the extensible DPC model works well out to
$f=400\,\pNunit$. Carrying the fit out to still larger values of $f$
would raise the fit value of $\Nestiff$ still further.

\section{The Overstretching Transition}
\subsection{Background}
As first observed by Cluzel {\em et
el.} \cite{Cluzel} and Smith {\em et al} \cite{Smith}, stretching
double stranded DNA is quite different from single-strand DNA. Their
experiments showed that at a force of around 65--70\,pN the DNA sample
suddenly snaps open (an ``overstretching transition''), extending
to almost twice its original contour
length before entering a second entropic stretching regime.
This second regime clearly represents a ``stretched'' DNA configuration quite
different from ordinary double stranded or B-DNA, and has been dubbed
S-DNA.  The transition
from B-DNA into S-DNA is very sharp, indicating a high level of
cooperativity.

S-DNA appears to have a definite helical
pitch \cite{lege99a,lege99b}, consistent with its being a new,
double-stranded conformation. An alternative view interprets the
overstretching transition as force-induced melting (denaturation) of
the B-DNA duplex \cite{RouzinaBloomfield1,RouzinaBloomfield2}. One
implication of the latter view is that S-DNA should have elastic
properties similar to those of two single strands, a point to
which we will return later.

Whatever view we take of its structural character, the sharpness of the
overstretching transition is reminiscent of another
well-studied structural transition in biopolymers, the helix-coil
transition \cite{ZimmBragg}. Inspired by the classic analysis of Zimm
and Bragg, this section will  model the B$\to$S transition by a
two-state (Ising) model living on a DPC (the ``Ising--DPC model''). We will
make no assumptions
about the nature of either B- or S-DNA. Both are allowed to have
arbitrary bend and stretch stiffnesses. Our
aim is to fit the resulting force-extension curves to the
available data and  to see whether the values
of the elastic constants can help characterize the stretched state.
(The other state is just double stranded DNA, whose elastic constants are well
known.)

\subsection{General Setup}
\begin{figure}
\begin{center}
  \includegraphics[width=.8\linewidth]{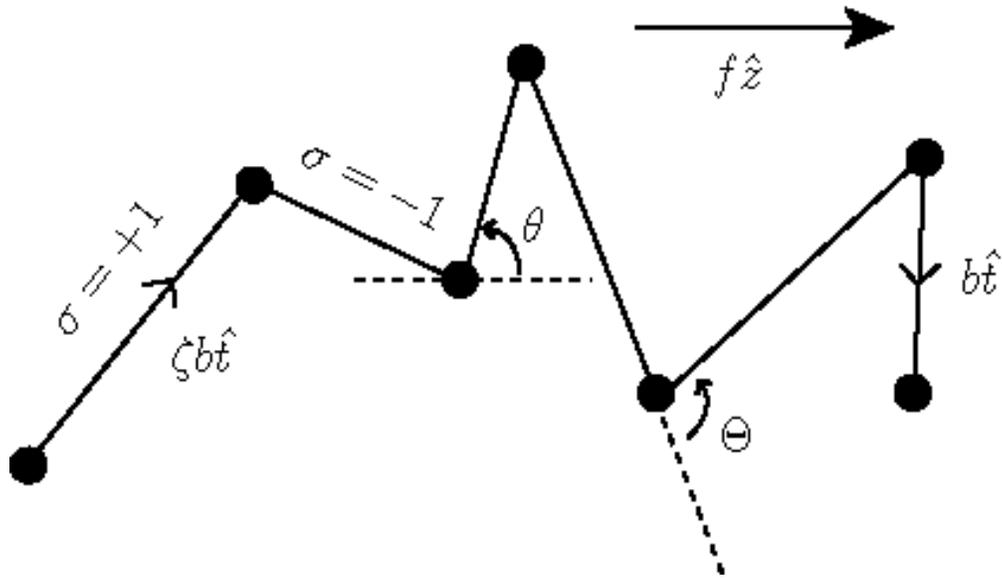}
\end{center}
\caption[]{Conventions for the Ising-DPC model. We take
$\sigma=+1$ to correspond to B-DNA, and $\sigma=-1$ to
S-DNA. Each segment of S-DNA is longer than B-DNA by a factor $\zeta$.
Definitions
of $\hat t, \theta$ and $\Theta$ are the same as before.}\label{isdpcfig}
\end{figure}

\fref{isdpcfig} illustrates the model that we will be
considering in some more detail. We envision a chain consisting of
N links, connected by hinges that try to align the segments they
join. Each segment carries a discrete variable $\sigma$, which takes
the values $\pm 1$. We will take $\sigma=+1$ to mean the segment is
in the B-state and $\sigma=-1$ for the S-state.  The factor by which a segment
elongates when going from B to S will be called $\zeta$, {\em
i.e.} $b^S=\zeta b$ (with $\zeta>1$). We assign a
bend stiffness parameter $A$ to B-DNA, and a different $A^{S}\equiv\beta
\zeta A$ to S-DNA; $\beta$ is a dimensionless parameter with $\beta\zeta<1$.
We also need to assign a bend stiffness to a
hinge joining a B and an S segment. This value we will call
$\Nxterm A$.

We can now write down the full energy functional for our Ising-DPC
model:
\begin{eqnarray}
\label{idpcen}
\frac{{\cal E}[\{\hat t_i,\sigma_i\}]}{\kb  T}&=&-\sum_{i=1}^{N-1} \biggl\{
\frac{\alpha_0}{2}(\sigma_i\!+\!\sigma_{i+\!1})+\gamma(\sigma_i
\sigma_{i+\!1}\!-\!1)\nonumber+
\\&&\hspace{-1.5cm}+\frac{fb}{2\kb
T}\left[(\frac{1\!+\!\sigma_i}{2}\!+\!\frac{1\!-\!\sigma_i}{2}\zeta)\hat
t_i \cdot \hat
z\!+\!(\frac{1\!+\!\sigma_{i+\!1}}{2}\!+\!\frac{1\!-\!\sigma_{i+\!1}}{2}\zeta)\hat t_{i+\!1}\cdot \hat z\right]\!- \nonumber \\
& &\hspace{-.6cm} -\frac{A}{2b}\left[
\frac{(1\!-\!\sigma_i)(1\!-\!\sigma_{i+\!1})}{4}\beta+|\sigma_i\!-\!\sigma_{i+\!
1}|\Nxterm+
\frac{(1\!+\!\sigma_i)(1\!+\!\sigma_{i+\!1})}{4}\right]
(\Theta_{i,i+\!1})^2 \biggr\} \, .
\end{eqnarray}
The first line is the pure-Ising part, with $2 \alpha_0 \kb T$ the
intrinsic free energy cost of converting a single segment from B to S and
$2\gamma \kb T$ the energy cost of creating
a B$\to$S interface. Note that we ignore a contribution to the energy
functional from the first and last segments. In the long-chain limit this
does not affect the outcome of our calculation.

The partition function for the energy functional (\ref{idpcen}), ${\cal
E}[\{ \hat t_i,\sigma_i \}]=\sum_{i=1}^{N-1}{\cal E}_i(\hat
t_i,\sigma_i,\hat t_{i+1},\sigma_{i+1})$, is given by
\begin{equation}
\label{2dpcZ}
{\cal Z}=\left[\prod_{i=1}^{N-1}\sum_{\sigma_i=\pm 1} \! \int_{{\mathbb
S}^2} \! {\rm d}^2\hat t_i\right] \!  \prod_{i=1}^{N-1}e^{-{\cal E}_i(\hat
t_i,\sigma_i,\hat t_{i+1},\sigma_{i+1})/\kb  T}\, .
 \end{equation}
We will again calculate ${\cal Z}$ with the aid of the transfer matrix
technique \cite{KramersWannier}, writing \eref{2dpcZ} as
\begin{equation}
{\cal Z}=\vec v \cdot {\sf T}^{N-1}\cdot \vec w\, ,
\end{equation}
with ${\sf T}$ now the transfer matrix for our Ising-DPC model,
which carries an additional 2-by-2 structure due to the Ising
variables. The dot products are thus defined as
\begin{equation}
({\sf T}\cdot \vec v)_{\sigma_i}(\hat t_i)=\sum_{\sigma_j=\pm 1} \!
\int_{{\mathbb S}^2} \! {\rm d}^2\hat t_j \, {\mathbb T}_{\sigma_i
\sigma_j}(\hat t_i,\hat t_j) v_{\sigma_j}(\hat t_j) \, .
\end{equation}
The individual matrix elements ${\mathbb T}_{\sigma_i \sigma_j}$ are given
explicitly by
\begin{eqnarray*}\label{Telts}
{\mathbb T}_{1,1}(\hat t_i,\hat t_{i+1})& = & \exp\left[\frac{1}{2} \tilde
f(\hat t_i\!+\!\hat
t_{i+1})\cdot \hat z\!-\!\frac{A}{b}(1\!-\!\hat t_i \cdot \hat
t_{i+1})+\alpha_0\right] \nonumber\\
{\mathbb T}_{1,\!-\!1}(\hat t_i,\hat t_{i+1})& = & \exp\left[\frac{1}{2}
\tilde f(\hat
t_i\!+\!\zeta\hat t_{i+1})\cdot \hat z\!-\!\frac{\Nxterm A}{b}(1\!-\!\hat
t_i \cdot
\hat  t_{i+1})\!-\!2 \gamma\right]\nonumber\\
{\mathbb T}_{-1,1}(\hat t_i,\hat t_{i+1})& = & \exp\left[\frac{1}{2} \tilde
f(\zeta\hat
t_i\!+\!\hat t_{i+1})\cdot \hat z\!-\!\frac{\Nxterm A}{b}(1\!-\!\hat t_i
\cdot \hat
t_{i+1})\!-\!2 \gamma\right]\nonumber\\
{\mathbb T}_{-1,-1}(\hat t_i,\hat t_{i+1})& = & \exp\left[\frac{1}{2} \zeta
\tilde f(\hat
t_i\!+\!\hat t_{i+1})\cdot \hat z\!-\!\frac{\beta A}{b}(1\!-\!\hat t_i
\cdot \hat
t_{i+1})\!-\!\alpha_0\right]\, ,
\end{eqnarray*}
where again $\tilde f\equiv\frac{fb}{\kb  T}$.

Once again we approximate the largest eigenvalue of the
transfer matrix ${\sf T}$  using a variational approach,
choosing our trial eigenfunctions to possess azimuthal symmetry
and to be peaked in the direction of the force $\hat z$. This
time, however, we need a three-parameter family of trial functions:
\begin{equation} \label{eomphi}
 v_{\omega_{1},\omega_{-1},\varphi}(\hat t\,)=
\left(\begin{array}{c } \left(\frac{\omega_{1}}{\sinh(2 \omega_1)}\right)^{\frac12} e^{\omega_1 \hat t \cdot \hat z} \cos \varphi \\
 \left(\frac{\omega_{-1}}{\sinh(2 \omega_{-1})}\right)^{\frac12}e^{\omega_{-1} \hat t \cdot \hat z}\sin \varphi \end{array}\right)\, ,
\end{equation}
chosen such that their squared norm is independent of all parameters
\begin{equation}
\| \vec v_{\omega_{1},\omega_{-1},\varphi}\|^2=2 \pi\, .
\end{equation}
\eref{eomphi} shows that once again the $\omega$'s gives the degree of
alignment of the monomers (how forward-peaked their probability
distribution is), whereas $\varphi$ describes the relative probability
of a monomer to be in the two states. The variational estimate for the
maximal eigenvalue is now given by
\begin{equation}\label{varesti}
\lambda^*_{\text{max}}\equiv\max_{\omega_{1},\omega_{-1},\varphi} y(\omega,\varphi)
\equiv\max_{\omega_{1},\omega_{-1},\varphi}\, \frac{\vec v_{\omega_{1},\omega_{-1},\varphi} \cdot
{\sf T} \cdot \vec v_{\omega_{1},\omega_{-1},\varphi}
}{\normsq{\vec v_{\omega_{1},\omega_{-1},\varphi}}} \, ,
\end{equation}

The maximization over $\varphi$ can be done analytically: defining the
$2\times 2$ matrix $\tilde {\sf T}(\omega_1,\omega_{-1})$ by
\begin{equation}
\vec v_{\omega_{1},\omega_{-1},\varphi} \cdot {\sf T} \cdot \vec
v_{\omega_{1},\omega_{-1},\varphi}=(\cos \varphi,\sin \varphi) \cdot \tilde {\sf
T}(\omega_1,\omega_{-1}) \cdot \left(\begin{array}{c }   \cos \varphi \\ \,
\sin \varphi \end{array}\right) \, ,
\end{equation}
or equivalently specifying its entries
\begin{equation}
\tilde {\mathbb T}_{\sigma_i \sigma_j}(\omega_{\sigma_i},\omega_{\sigma_j})=\int_{{\mathbb S}^2} \! {\rm
d}^2 \hat t_i\int_{{\mathbb S}^2} \! {\rm d}^2\hat t_{j}
\,\left(\frac{\omega_{\sigma_i}}{\sinh(2 \omega_{\sigma_i})}\right)^{\frac12} e^{\omega_{\sigma_i} \hat t_i \cdot \hat z}\,{\mathbb T}_{\sigma_i \sigma_j}(\hat
t_i,\hat t_j)\left(\frac{\omega_{\sigma_j}}{\sinh(2 \omega_{\sigma_j})}\right)^{\frac12} e^{\omega_{\sigma_j} \hat t_j \cdot \hat z}\, ,
\end{equation}
it is easy to show that
\begin{equation}
\lambda^*_{\text{max}}=\max_{\omega_1,\omega_{-1}}\, \frac{\tilde y(\omega_1,\omega_{-1})}{\normsq{\vec
v_{\omega_{1},\omega_{-1},\varphi}}}\, ,
\end{equation}
where $\tilde y(\omega_1,\omega_{-1})$ is the maximal eigenvalue of
$\tilde{\sf T}(\omega_1,\omega_{-1})$. The following subsection will calculate this
eigenvalue in a
continuum approximation to $\tilde{\sf T}(\omega_1,\omega_{-1})$,  illustrating the
procedure by considering in some detail the matrix element $\tilde{\mathbb
T}_{1,1}(\omega_1,\omega_{-1})$.  The other matrix elements can be obtained analogously.
Writing out the integrals explicitly, we have
\begin{equation}
\tilde{\mathbb T}_{1,1}(\omega_1)=\frac{\omega_{1} e^{\alpha_0-\frac{A}{b}}}{\sinh(2 \omega_1)}\int_{{\mathbb
S}^2} \! {\rm
d}^2\hat t_i e^{\hat a \hat t_i \cdot \hat z}\int_{{\mathbb S}^2} \! {\rm
d}^2\hat t_{i+1}
\left[e^{(\hat a \hat z+\frac{A}{b}\hat t_i)\cdot \hat t_{i+1}}\right]\, ,
\end{equation}
where we have introduced $\hat a \equiv \omega_1+\frac{\tilde f}{2}$.
Condensing notation even further we define $\mu^2=\hat
a^2+(\tfrac{A}{b})^2+2\hat a\tfrac{A}{b}\hat t_i \cdot \hat z$, which
allows us to write
\begin{equation}\label{eqT}
\tilde{\mathbb T}_{1,1}(\omega_1)\!=\!(2
\pi)^2\frac{\omega_{1} e^{\alpha_0-\frac{A}{b}}}{\sinh(2 \omega_1)}\int_{|\tfrac{A}{b}\!-\!\hat a|}^{\tfrac{A}{b}\!+\hat a}\!\!  \
\frac{ b\,{\rm d}\mu}{\hat a A} e^{\frac{b}{2 A}(\mu^2\!-\!\hat
a^2\!-\!(\frac{A}{b})^2)}\left[ e^\mu\!-\!e^{-\mu}\right]\, .
\end{equation}

\subsection{Continuum Limit\label{s:cl}}
We could now proceed to evaluate the force-extension relation of the
Ising-DPC model, by generalizing \sref{secdpc}. To simplify the
calculations, however, we will first pass to a continuum limit. To
justify this step, note that \fref{compareall} shows that the
continuum (WLC) approximation gives an excellent account of
single-stranded DNA stretching out to forces beyond those probed
in overstretching experiments (about $90\,\pNunit$).  As mentioned
earlier, the continuum approximation is also quite good for double-stranded
DNA, because the latter's persistence length is so much longer than
its monomer size.

In the continuum limit $b$ is sent to zero holding $L\tot$ fixed; hence
$N\to\infty$. The bookkeeping is more manageable after a shift in
$\mu$:
\begin{equation}
x\equiv \mu-\frac{A}{b}\, .
\end{equation}
\eref{eqT} then reduces to
\begin{eqnarray}
\tilde{\mathbb T}_{1,1}(\omega_1)\!&=&\!\frac{\omega_{1} e^{\alpha_0}}{\sinh(2 \omega_1)} \frac{(2 \pi)^2 b}{\hat a
A}\int_{-\hat a}^{+\hat a} \!\!  \, {\rm d}x\, \exp \left[ \frac{b}{2 A}
x^2 +2x-\frac{\hat a^2 b}{2A}\right] \nonumber \\
&\approx& \!\frac{\omega_{1} e^{\alpha_0}}{\sinh(2 \omega_1)} \frac{(2 \pi)^2
b}{\hat a A}\int_{-\hat a}^{+\hat a} \!\!  \, {\rm d}x\, e^{2x}(1+\frac{x^2
b}{2 A}) e^{-\frac{\hat a^2 b}{2 A}}\, .
\end{eqnarray}
The last integral can be worked out exactly, and expanding the result to second order in $b$ we end up with 
\begin{equation}
\frac{A}{2\pi b} \frac{1}{\normsq{\vec v_{\omega_{1},\omega_{-1},\varphi}}}\tilde{\mathbb
T}_{1,1}(\omega_1)\!=\!e^{\alpha_0}\left[ 1+{b}\left(\frac{f}{\kb T}-\frac{\omega_1}{2A}\right)\left(\coth(2\omega_1)-\frac{1}{2\omega_1}\right)\right]\, .
\end{equation}
In similar fashion, we can obtain the following expressions for the other
matrix elements.
\begin{eqnarray}
\frac{A}{2\pi b} \frac{1}{\|\vec v_{\omega_{1},\omega_{-1},\varphi}\|^2}\tilde{\mathbb
T}_{-1,-1}(\omega_{-1})\!&=&\!\beta\inv e^{-\alpha_0}\left[
1+{b}\left(\frac{\zeta
f}{\kb  T}-\frac{\omega_{-1}}{2 \beta
A}\right)\left(\coth(2\omega_{-1})-\frac{1}{2\omega_{-1}}\right)\right] \nonumber \\
\frac{A}{2\pi b} \frac{1}{\|\vec v_{\omega_{1},\omega_{-1},\varphi}\|^2}\tilde{\mathbb
T}_{1,-1}(\omega_1,\omega_{-1})\!&=&\!\frac{e^{-2\gamma}}{\Nxterm} \left( \frac{\omega_1 \omega_{-1}}{\sinh(2\omega_1)\sinh(2\omega_{-1})}\right)^{\frac12} \left( \frac{2 \sinh(\omega_1+\omega_{-1})}{\omega_1+\omega_{-1}}\right)\,.
\end{eqnarray}

To obtain a nontrivial continuum limit we must now specify how the
parameters $A$, $\alpha_0$, and $\gamma$ depend on $b$ as $b \to 0$. It is
straightforward to show that the choices
\begin{equation}
\alpha_0=-\frac12\ln\beta+b\bar\alpha\,,\qquad \gamma=-\frac12\ln(\bar
g b)
\end{equation}
work, where we hold $A$, $\bar\alpha$, $\beta$ and $\bar g$ fixed as $b\to0$.
With these choices, the matrix
$\frac1{\normsq{\vec v_{\omega_{1},\omega_{-1},\varphi}}}\tilde{\sf T}(\omega_{1},\omega_{-1})$ takes the form
\begin{equation}
\label{tmatcons}
 \frac{1}{\|\vec v_{\omega_{1},\omega_{-1},\varphi}\|^2}\tilde{\sf T}(\omega_{1},\omega_{-1})=\frac{2 \pi
b}{A \sqrt{\beta}}\left(\openone +b
\begin{pmatrix}
    {\cal P}  & {\cal Q}    \\
 {\cal Q}     & {\cal R}
\end{pmatrix}\right)\, ,
\end{equation}
with
\begin{eqnarray}
{\cal P}&=&\bar\alpha  +
\left(\frac{f}{\kb
T}-\frac{\omega_1}{2A}\right)\left(\coth(2\omega_1)-\frac{1}{2\omega_1}\right)\nonumber\,
,\\
{\cal R}&=&-\bar\alpha  + \left(\frac{\zeta
f}{\kb T}-\frac{\omega_{-1}}{2A\beta
}\right)\left(\coth(2\omega_{-1})-\frac{1}{2\omega_{-1}}\right)\nonumber \, ,\\
{\cal Q}&=&\frac{\bar g \sqrt{\beta}}{\Nxterm}\left( \frac{\omega_1 \omega_{-1}}{\sinh(2\omega_1)\sinh(2\omega_{-1})}\right)^{\frac12} \left( \frac{2 \sinh(\omega_1+\omega_{-1})}{\omega_1+\omega_{-1}}\right)\, .
\end{eqnarray}
Note that the prefactor $\frac{2\pi b}{A \sqrt\beta}$ in \eref{tmatcons}
does not
contribute to the force-extension result \eref{zldef}, since it
does not depend on the force.  In terms of the individual matrix
entries, the quantity to be maximized now reads (see \eref{varesti}):
\begin{equation}
\label{lamom}
\ln \tilde y(\omega_{1},\omega_{-1})=\frac{b}{2}\left({\cal P}+{\cal
R}+\sqrt{({\cal P}-{\cal R})^2+4{\cal Q}^2}\right)\, .
\end{equation}
Writing $\Omega\equiv b\inv\ln\lambda^*_{\text{max}}=
b\inv\times\max_{\omega} \ln
\tilde y(\omega_{1},\omega_{-1})$, the force-extension in the continuum limit is finally
given by
\begin{equation}
\langle \frac{z}{L_{\rm tot,b}}\rangle=\kb  T\frac{{\rm d}\Omega}{{\rm d}
f} \, .
\end{equation}
We evaluate $\Omega$ by numerically maximizing \eref{lamom}.

So far, we have not included stretch moduli for the B- and
S-DNA. This is easily implemented to first order in $f/\Nestiff$ by
replacing $f$
with $f(1+\frac{f}{2 \Nestiff^{S,B}})$ in the matrix elements for the two
states respectively (\eref{Telts}). This procedure yields theoretical
force-extension curves like the ones plotted in Figs.~(\ref{smithfit}) and
(\ref{cluzelfit}).

In summary, our model contains the following seven parameters.  $2
\bar \alpha \kb T$ is the free energy per unit length required to flip B-DNA into
the $S$-state, and is measured in [J/nm].  ${\cal Q}$ measures
the cooperativity of the transition and has units [1/nm].  $A$ is the
bend stiffness parameter of B-DNA, with units [nm].  The dimensionless parameter $\beta$
is the ratio of the B- and S-DNA bend stiffnesses.  $\Nestiff^B$ and
$\Nestiff^S$ are the stretch stiffnesses of B and S-DNA, and are
measured in pN. Finally, $\zeta$ is the dimensionless elongation
factor associated with the B$\to$S transition.
\subsection{Discussion of fits\label{fitdisc}}
Our strategy is now as follows: first, we fit the part of the stretching curve well below 65\thinspace pN to a one-state,
continuum model (\textit{i.e.} to the EWLC), determining its effective
spring constant and stretch modulus. The values thus obtained are used
as initial guesses in a fit of the full curve to the Ising-DPC model. To
improve convergence, we eliminate two of the
parameters, as follows. First, we can get an accurate value for
$\Nestiff^B$ from the low force data, so we hold it  fixed to this
value during the full fit. Second,
as described in \sref{secdpc} we can work out
the low-force limit analytically, and from this obtain the effective
spring constant $\Neffsp$ as a function of the model's parameters.
We invert this relation to get $A$ as a function of $\Neffsp$ and the other
parameters. We substitute this $A$, holding $\Neffsp$ fixed to
the value obtained by fitting the low-force data to an EWLC. We then fit the
remaining five parameters ($\beta$, $\cal Q$, $\bar\alpha $, $\Nestiff^{S}$
and $\zeta$) to the dataset\footnote{In our fits, we exclude the data points in the steepest region of the graph. Because of the inevitable scatter in the data and the fact that only the deviations in the $y$-direction enter into $\chi^2$ their residuals are overemphasized, hindering convergence and accuracy of the routine.}.

\begin{figure}
\begin{center}
  \includegraphics[width=.8\linewidth]{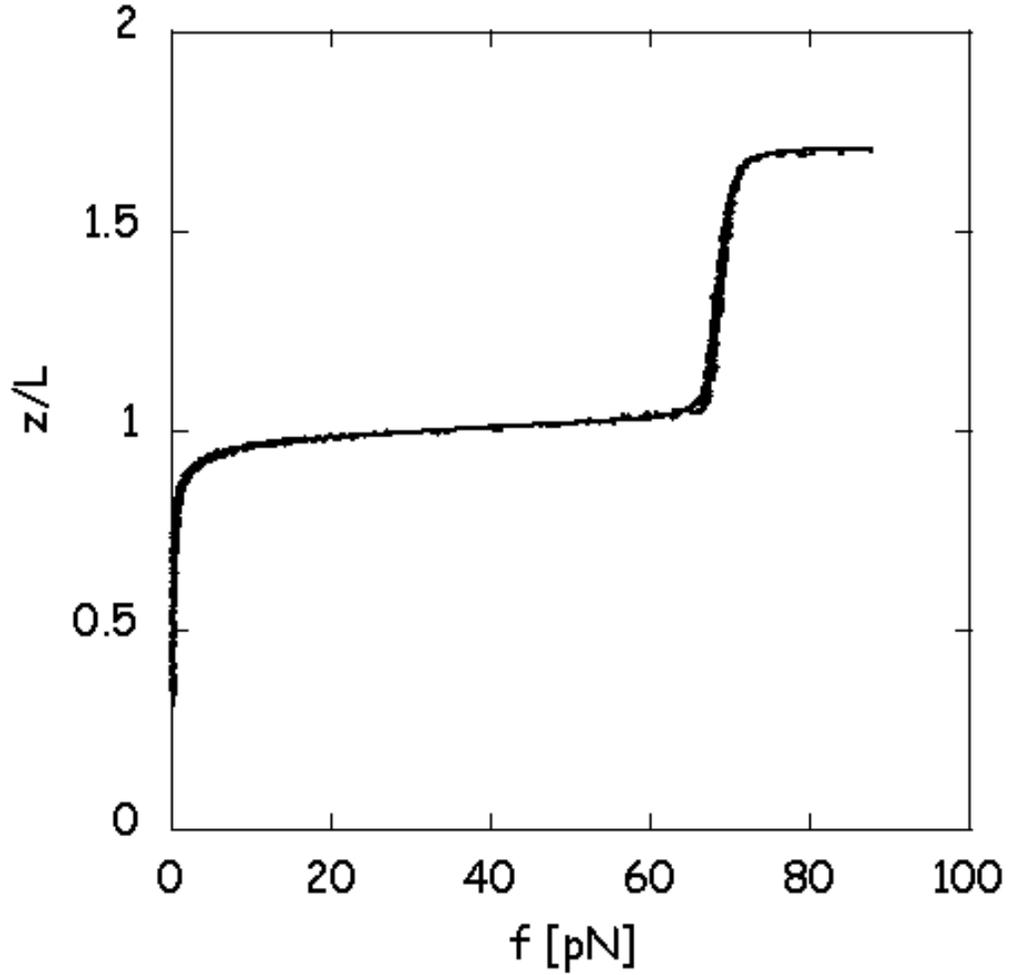}
\end{center}
\caption[]{Least-squares fit of the Ising-DPC model to an overstretching
dataset  (48.5\thinspace kbp $\lambda$
DNA construct; buffer 500\thinspace mM NaCl,
20\thinspace mM Tris, pH~8). Data kindly supplied by C.~Bustamante and
S.~Smith.
Fit parameters: $\Neffsp^{\dpc}=\frac{3 \kb
T}{2}\frac{1}{43.75\thinspace{\rm nm}},
\bar\alpha =5.45\, , \beta=0.16\, ,{\cal Q}=0.13\, ,\zeta=1.76\, ,
\Nestiff^{B}=1.2\cdot 10^3\,\pNunit$ and
$\Nestiff^{S}=1.0\cdot10^4\,\pNunit$. $\chi^2=9.22$ at $N=825$, points with $1.11<\langle\frac{z}{L}\rangle < 1.55$ were excluded from the fit. For further
discussion see \sref{fitdisc}.}\label{smithfit}
\end{figure}

\begin{figure}
\begin{center}
  \includegraphics[width=.8\linewidth]{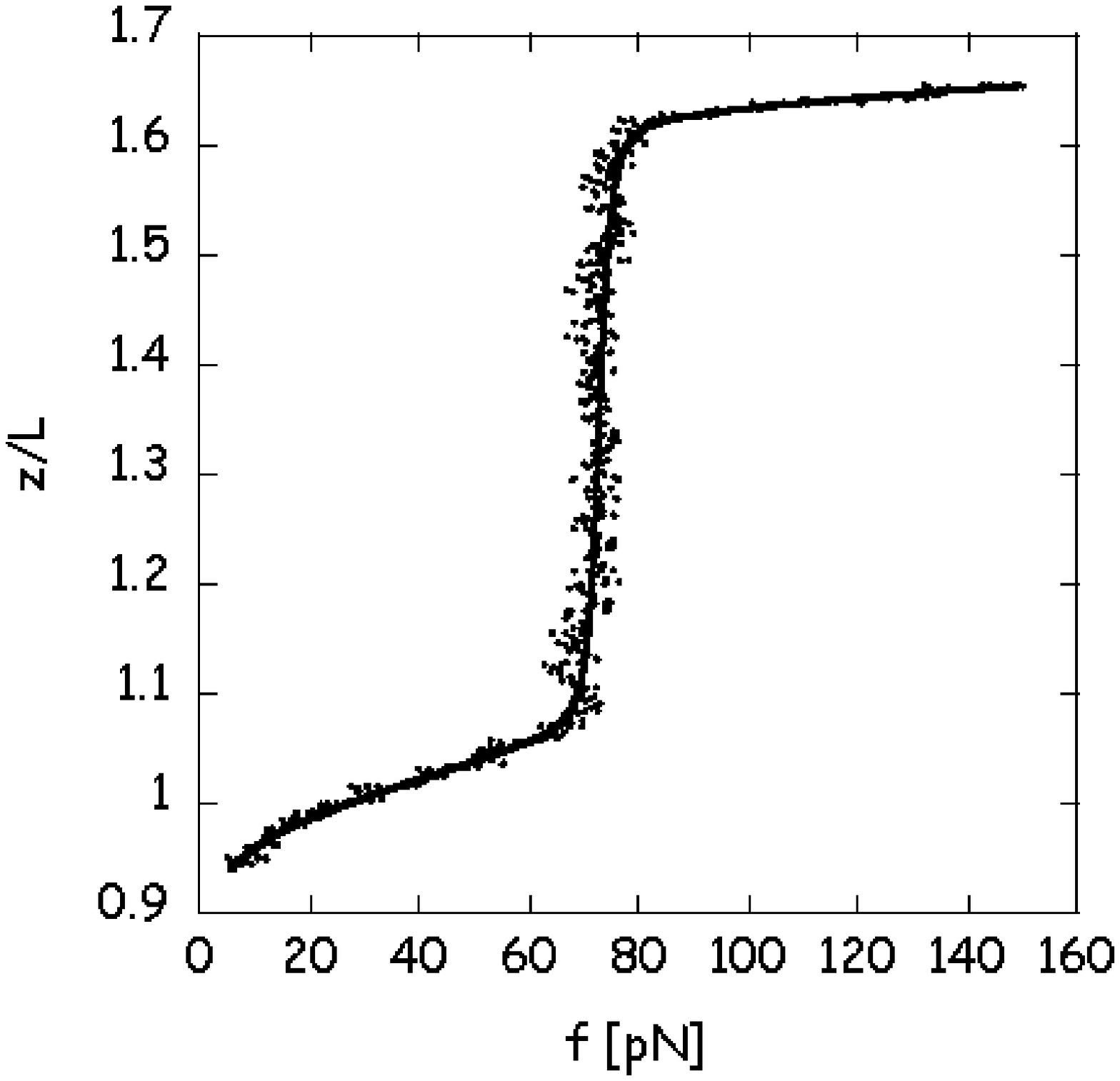}
\end{center}
\caption[]{Least-squares fit of the Ising-DPC model to an overstretching
dataset obtained from a 15.1 $\mu$m sample of EMBL3 $\lambda$ DNA in phosphate-buffered solution (100mM; 80mM Na$^+$ and 0.01\% Tween) from  \cite{Cluzel}. Data kindly supplied by
J. Marko. Fit parameters: $\Neffsp^{\dpc}=\frac{3 \kb
T}{2}\frac{1}{52.63\thinspace{\rm nm}}, \bar
\alpha_0=4.82\,\mathrm{nm}\inv\, , \beta=0.08\, ,{\cal Q}=0.23\,
,\zeta=1.71\, ,
\Nestiff^{B}=7.3\cdot 10^2\,\pNunit$ and
$\Nestiff^{S}=3\cdot10^4\,\pNunit$. $\chi^2=2.15$ at $N=339$, points with $1.15<\langle\frac{z}{L}\rangle < 1.5$ were excluded from the fit. For
further discussion see \sref{fitdisc}. }\label{cluzelfit}
\end{figure}

The results of the fits obtained in this manner are collected in
Figs.~(\ref{smithfit}) and (\ref{cluzelfit}).  Our Ising-DPC hybrid model fits
the experimental data rather well, but with so many fit parameters one
may ask whether the model actually makes any falsifiable predictions.
To answer this question we note that the data below the transition
suffice to fix $A$ and $\Nestiff^B$ as usual, roughly speaking from the
curvature and slope of the curve below the transition. Similarly,
the data above the transition fix $A^{S}=\zeta\beta A$ and
$\Nestiff^S$. The vertical jump in the curve at the transition fixes
$\zeta$. The horizontal location of the jump fixes $\bar\alpha$, and
the steepness of the jump fixes the cooperativity ${\cal
Q}$.\footnote{The fit value of $\bar\alpha$ should be regarded as an
average of the two different costs to convert AT or GC pairs. The fit
value of ${\cal Q}$ has no direct microscopic significance, as the
apparent cooperativity of the transition will be reduced by the
sequence disorder.} Thus all
of the model's parameters are fixed by specific features of the data.
Two additional, independent features of the data now remain, namely the
rounding of the curve at the start and end of the transition. Our
model predicts these features fairly succesfully.

Some common features emerging from
the two fits deserve comment.  First, both fits reproduce the
known values for the effective persistence length of B-DNA of around
$50\,$nm and its stretch modulus of about $1000\,$pN. Second, we
can read off the bend stiffness of S-DNA from our fit as
$A^{S}=\beta\zeta A=12.32\,$nm (data from \fref{smithfit}) or 7.2\thinspace nm
(data from \fref{cluzelfit}). If S-DNA consisted of two unbound, single
strands, we might have expected $A^S$ to be twice as large as the value
$A^{\rm ss}\approx0.85\,$nm obtained by fitting the single-strand stretching
data
with the continuum EDPC model (see \fref{compareall}). On the contrary, we
find that  {\em the bend stiffness of S-DNA is intermediate between that of
B-DNA and that of
two single strands}. This conclusion fits qualitatively with some of
the structural models of S-DNA, in which the bases remain paired but
are not stacked as in B-DNA.

Our third conclusion is that {\em the stretch modulus of S-DNA is
substantially higher than that of B-DNA.} This conclusion is again consistent with the view of S-DNA as stabilized mainly by its
backbones, which are much straighter than in B-DNA; the contour length
of B-DNA is instead determined by weaker, base-stacking interactions.

\subsection{Relation to prior work}
Polymer models with both finite cutoff and steric hindrances to motion
are not new. Classical examples include the rotation-isomer models, in
which succeeding monomers are joined by bonds of fixed polar angle but
variable azimuthal angle \cite{Grosberg}. Models of this sort have had
some success in making a priori predictions of the persistence length
of a polymer from its structural information, but obtaining the
force-extension relation is mathematically very difficult. Thus for
example \cite{miya62} obtain only the first subleading term in the
low-force expansion. We are not aware of a prior formulation of a
model incorporating the microscopic physics of discreteness and
stiffness, with a detailed experimental test.

Several authors have also studied the entropic elasticity of
two-state chains. As soon as the overstretching transition was
discovered, Cluzel proposed a pure Ising model by analogy to the
helix-coil transition \cite{cluz96a}. Others then introduced entropic
elasticity, but required that both states have the same bending
stiffness as B-DNA \cite{mark98a,ahsa98} or took one of the
two states to be
infinitely stiff \cite{tama01}, or to be a FJC
\cite{RouzinaBloomfield1,RouzinaBloomfield2}. We believe our Ising-DPC
model to be the first consistent formulation incorporating the
coexistence of two different  states with arbitrary elastic constants.
Our approach also is calculationally  more straightforward than some,
and minimal in the sense that no unknown potential function needs to
be chosen.

\section{Statistical analysis of the B$\to$S transition}
Using standard techniques from statistical physics, we now look at
the B$\to$S transition in some more detail.  From the expressions for
the Ising-DPC hybrid energy functional (\ref{idpcen}) and the
partition function (\ref{2dpcZ}) we read off that the average ``spin"
$\sigma$ can be obtained as
\begin{equation}
\langle \sigma \rangle=\frac{1}{N}\frac{\partial}{\partial \alpha_0} \ln
{\cal Z}=\frac{\partial}{\partial \bar\alpha }\Omega \, ,
\end{equation}
so that for instance the relative population of the S-state (or
equivalently the probability to find an arbitrary segment in the S-state),
$P(S)$, is given by
\begin{equation}
\label{pofs}
P(S)=\frac{1}{2}(1-\langle \sigma \rangle) \, .
\end{equation}
Similarly, we can take the derivative of \eref{2dpcZ} with respect to
$\gamma$ to determine the average nearest neighbor spin correlator
\begin{equation}
\langle \sigma_i \sigma_{i+1} \rangle=\frac{1}{N}\frac{\partial}{\partial
\gamma} \ln {\cal Z}+1=1-2b {\cal Q}\frac{\partial}{\partial  {\cal
Q}}\Omega\, .
\end{equation}
The quantity $\langle \sigma_i \sigma_{i+1} \rangle$ can be interpreted as the
fraction of nearest neighbor pairs in the same state minus the
fraction of pairs in  opposite states.  Consequently, the probability
of having a spin flip at a given site is $P(\text{flip})=\tfrac{1}{2}(1-\langle
\sigma_i \sigma_{i+1} \rangle)$ and the average {\em number} of S+B
domain pairs is $N_{\mathrm{pairs}}=\tfrac{N}{2}P(\text{flip})$.  A
heuristic measure of the typical S-domain size is then \cite{cantorbook}
\begin{equation}
\label{loff}
L_{\mathrm{dom}}=\frac L{N_{\rm pairs}}P(S)=
\frac{2b(1-\langle \sigma \rangle)}{1-\langle \sigma_i \sigma_{i+1} \rangle}
=\left(1-\frac{\partial\Omega}{\partial\bar\alpha}\right)/
\left({\cal Q}\frac{\partial\Omega}{\partial{\cal Q}}\right)
\,.\end{equation}

\begin{figure}
\begin{center}
  \includegraphics[width=.8\linewidth]{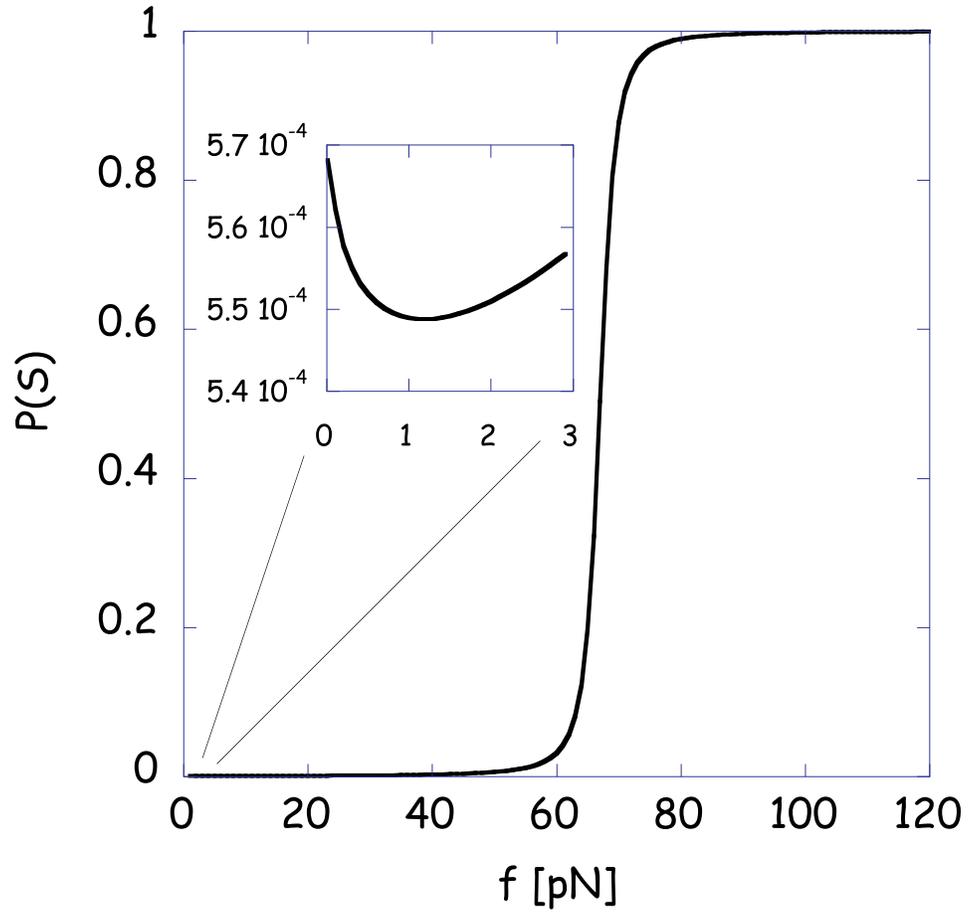}
\end{center}
\caption[]{$P(S)$, the relative population of the S-state, versus the applied
stretching force, as calculated from \eref{pofs}. The inset shows that
the S-state has a nonzero population even at zero force. Parameter
values are those from Fig. \ref{cluzelfit}.}\label{sigav}
\end{figure}

\begin{figure}
\begin{center}
  \includegraphics[width=.8\linewidth]{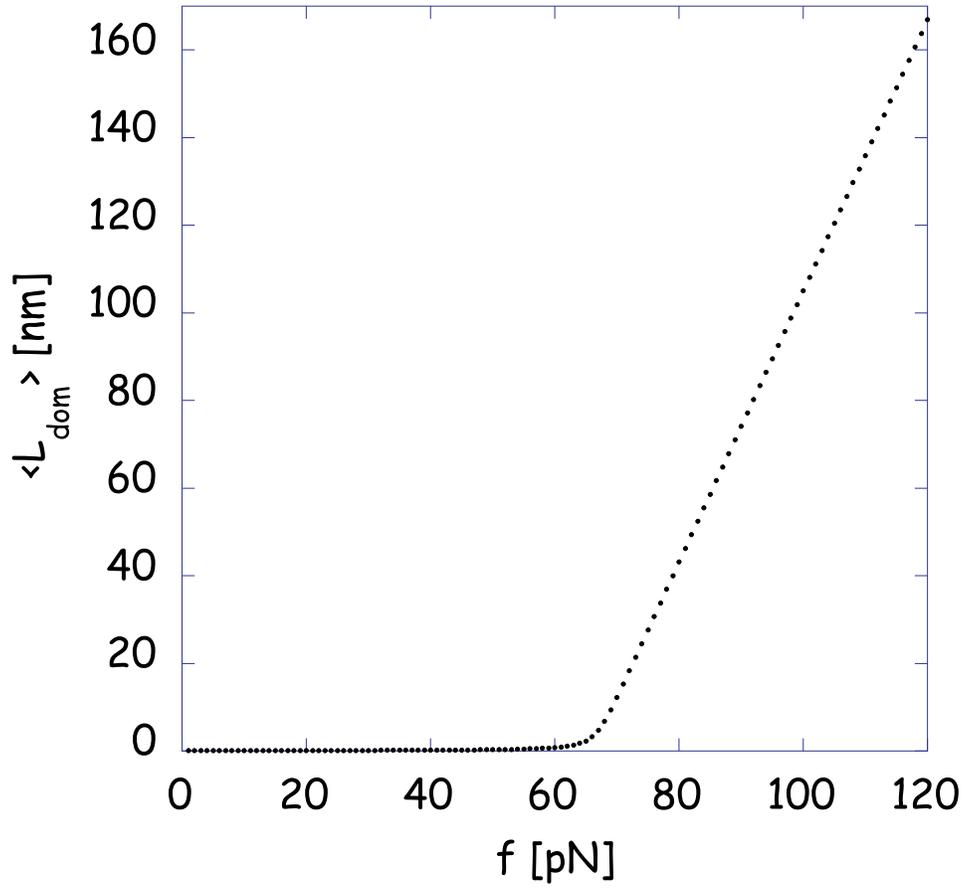}
\end{center}
\caption[]{The typical length of an S-domain $L_{\mathrm{dom}}$ vs. the
stretching force, calculated using \eref{loff}. Parameter values are
those of Fig. \ref{cluzelfit}. The asymptotic slope of the
linear increase has been determined to be $3.15$nm pN$^{-1}$. Note, that
even at $120$pN, the typical size of an S-domain is only 160nm, or about
480 basepairs.}\label{ldom}
\end{figure}

We wish to highlight two points from this discussion. First,
\fref{sigav} shows the fraction in the S-state, $P(S)$, as a
function of the applied force, and we can see the characteristic
sigmoidal behavior as the system is led through the transition.  As
the inset demonstrates, a small fraction is in the S-state even at
zero force.  This fraction initially decreases upon increasing the
stretching force.\footnote{A related reentrant phenomenon was noted in
\cite{tama01}.}
\fref{ldom} plots the typical S-domain length
$L_{\rm dom}$ versus applied stretching force.  It demonstrates how even
well above the transition
the S-state on average does not persist for very long; at the high end
of the physically accessible range of forces S-domains measure about
160nm.  This figure has some significance as it illustrates an
important point about the role of nicks in the experiments.  Empirically,
when working with $\lambda$-phage DNA only around
5\% of all samples are completely unnicked \cite{lege99b}.  Since the
$\lambda$-phage
genome is about 48Kbp in length, we can roughly estimate the
probability for an arbitrary base pair to be unnicked is
$P(\text{not})=(0.05)^{1/48000}$, and consequently the probability
that a given pair {\em is} nicked is $P(\text{nick})=1-P(\text{not})
\approx 6.2\cdot 10^{-5}$.  Given the total length of $\lambda$-phage
DNA, this implies we expect there to be an average of $6.2\cdot
10^{-5}\times 48\cdot 10^3\approx 3$ nicks per sample, corresponding
to an average distance between nicks of the order of 5$\,\mu$m,
considerably larger than the typical S-domain size. This observation
bears on the question of the character of the S state of DNA
\cite{RouzinaBloomfield1}:   even if S-DNA were a denatured state, the
existence of nicks would not necessarily cause it to
suffer irreversible changes in its elasticity as tracts spanning two
nicks fall off during overstretching.

Secondly, different groups have not agreed on whether the stretching curves of
double-stranded and single-stranded DNA coincide at forces above the
former's overstretching transition \cite{bust01a,lege99b}. We wish to
point out that even if S-DNA were a denatured state, we still would not
necessarily expect these two curves to coincide. \fref{sigav} shows that
the conversion from
B- to S-form continues well beyond the apparent end of the force
plateau, continuing to affect the force-extension curve. To determine
whether S-DNA is elastically similar to B-DNA one must disentangle the
two states' contributions to the stretching curve by globally fitting
to a 2-state model, as we have done.

\section{Conclusion}
\sref{s:IS} summarizes our conclusions. Here we list a number of
interesting modifications to the model, as possible extensions
to this work.

While the variational approximation used here has proved to be
adequate, still it is straightforward to replace it by the
eigenfunction-expansion technique, which can be carried to arbitrary
accuracy \cite{MarkoSiggia}. Similarly, the methods of \sref{secdpc}
can be used to work in the full, discrete DPC model instead of the
continuum approximation used in \sref{s:cl}. It is also straightforward
to retain finite-length effects, by keeping the subleading eigenvalue
of the transfer matrix.

Real DNA is not a homogeneous rod. The methods of quenched disorder can
be used to introduce sequence-dependent contributions to the
transition free energy $\alpha$ and the bend stiffness $A$. Finally, we
believe that the methods of this paper can be adapted to the study of
the stretching of individual polypeptide and polysaccharide molecules
\cite{rief98}.

\begin{acknowledgments}
We thank T. Burkhardt, D. Chatenay, A. Grosberg, R. Kamien, J. Marko and M. Rief and for valuable
discussions, and  C.~Bustamante, D.~Chatenay, J.-F. L\'eger, J. Marko, M.
Rief, and S. Smith for sending us experimental data.
CS acknowledges support from NIH grant R01 HL67286 and from NSF grant DMR00-79909. 
PN acknowledges support from
from NSF grant DMR98-07156.
\end{acknowledgments}

\appendix
\section{Derivation of $y(\omega)$, the variational approximation to
$\lambda_{\text{max}}$.}\label{appx}
In this appendix we will derive an expression for $\Nlaest(\omega)$ as
defined in \eref{laestdef}, which we recall reads
\begin{equation}\label{la}
\normsq{\vec v_\omega}\Nlaest(\omega)\equiv \vec v_\omega \cdot {\sf T} \cdot
\vec v_\omega \, .
\end{equation}
We will assume that the angles $\Theta_{i,i+1}$ between successive links
are small, which allows us to replace $(\Theta_{i,i+1})^2=\arccos^2(\hat
t_i \cdot \hat t_{i+1})$ by its small-angle approximation $2(1-\hat t_i
\cdot \hat t_{i+1})$. The family of trial functions we use is parameterized
by the single parameter $\omega$; $v_\omega(\hat t\,)\equiv e^{\omega \hat t
\cdot \hat z}$. Furthermore, we will ignore the two contributions from the
beginning and end of the chain (appearing for instance in \eref{pfdpc}),
as they do not contribute to our result in the long chain limit anyway. Thus
the energy functional is
\begin{equation}
\frac{{\cal E}[\{\hat t_i\}]}{\kb  T}=-\sum_{i=1}^{N-1}\left\{\frac{fb}{2 \kb
T} \, (\hat t_i \cdot \hat z+\hat t_{i+1} \cdot \hat z)\,
-\frac{A}{b}(1-\hat t_i \cdot \hat t_{i+1}) \right\} \, .
\end{equation}
According to \eref{opent}, the matrix elements of ${\sf T}$ are given by
\begin{equation}
{\mathbb T}(\hat t_i,\hat t_{i+1})=\exp\left[ -\tilde \ell +\frac{\tilde
f}{2}(\hat t_i+\hat t_{i+1}) \cdot \hat z+\tilde \ell \,\hat t_i \cdot \hat
t_{i+1}\right]\, ,
\end{equation}
where we use the dimensionless force $\tilde f\equiv \frac{f b}{\kb T}$ and
ratio of characteristic lengths $\tilde \ell=\frac{A}{b}$.
Working out the scalar products in \eref{la} yields
\begin{equation}\label{bigla}
\normsq{\vec v_\omega}\Nlaest(\omega)=e^{-\tilde \ell}\int_{{\mathbb S}^2} \! {\rm
d}^2\hat t_i \int_{{\mathbb S}^2} \! {\rm
d}^2\hat t_{i+1}
\exp\left[(\frac{\tilde f}{2}+\omega)(\hat t_i+\hat t_{i+1}) \cdot \hat
z+\tilde \ell \,\hat t_i \cdot \hat t_{i+1}\right] \, .
\end{equation}
Defining an auxiliary vector
\begin{equation}
\vec G\equiv (\frac{\tilde f}{2}+\omega)\hat z+\tilde \ell \,\hat t_i
\equiv G\,\hat g \, ,
\end{equation}
with
\begin{equation}
G \equiv \|\vec G\|=\left( (\frac{\tilde f}{2}+\omega)^2+\tilde \ell ^2 +
\tilde\ell (\tilde f +2\omega)\,\hat t_i \cdot \hat z \right)^{\frac12}\, ,
\end{equation}
simplifies \eref{bigla}, which now reads
\begin{equation}
\normsq{\vec v_\omega}\Nlaest(\omega)=e^{-\tilde \ell}\int_{{\mathbb S}^2} \!
{\rm d}^2\hat t_i \exp\left[  (\frac{\tilde f}{2}+\omega)\hat t_i \cdot
\hat z \right]  \int_{{\mathbb S}^2} \! {\rm
d}^2\hat t_{i+1} \exp \left[ G \, \hat g\cdot \hat t_{i+1}\right]\, .
\end{equation}
Transforming to spherical polar coordinates with $\hat g$ as the polar
axis, the second integral can be worked out to give $\frac{4
\pi}{G}\sinh(G)$. Since the integral over $\hat t_i$ involves only terms
containing $\hat t_i \cdot \hat z$, the integration over the azimuthal
angle simply yields $2 \pi$. For the polar angle, we change the integration
variable to $G$ (which is a monotonic function of  $\hat t_i \cdot
\hat z$), bringing it to the following form
\begin{equation}
\normsq{\vec v_\omega}\Nlaest(\omega)=\frac{16\pi^2}{\tilde \ell (\tilde
f+2\omega)}\exp\left[-\frac32 \tilde \ell-\frac{1}{2 \tilde
\ell}\left(\frac{\tilde f}{2}+\omega\right)^2\right]\int_{|\tilde
\ell-(\frac{\tilde f}{2}+\omega)|}^{\tilde \ell+(\frac{\tilde
f}{2}+\omega)}\! {\rm d} G\, \exp\left[G^2/2 \tilde \ell\right]\sinh(G)\, .
\end{equation}
The integral over $G$ can be performed analytically, and is most conveniently expressed in terms of error functions as
\begin{eqnarray}
\int_{|\tilde \ell-(\frac{\tilde f}{2}+\omega)|}^{\tilde \ell+(\frac{\tilde
f}{2}+\omega)}\! {\rm d} G\, \exp\left[G^2/2 \tilde
\ell\right]\sinh(G)&&=\nonumber \\
&& \hspace{-5cm} \frac{e^{-\tilde \ell/2} \sqrt{-\pi \tilde \ell}}{2 \sqrt
2} \left[ \text{Erf}\left( \frac{i}{2\sqrt{2 \tilde \ell}} (\tilde f + 4
\tilde \ell + 2 \omega)\right)-\text{Erf}\left( \frac{i}{2\sqrt{2 \tilde
\ell}} (\tilde f - 4 \tilde \ell + 2 \omega) \right)\right] \, .
\end{eqnarray}
This expression is valid only in the regime where $\tilde \ell>\frac{\tilde
f}{2}+\omega$, which is satisfied as long as one chooses $A>b$. Note that
the error functions have imaginary arguments. Using the normalization
quoted in \eref{vnorm} we can now express $\Nlaest(\omega)$ in a form that
is well suited for further (numerical) manipulations:
\begin{eqnarray}
\Nlaest(\omega)& = & \frac{2 \sqrt 2 \pi^{3/2} \omega e^{-2 \tilde \ell
-\frac{(2\omega+\tilde f)^2}{8\tilde \ell}}\text{csch}(2
\omega)}{\sqrt{-\tilde \ell}(2\omega+\tilde f)} \times \nonumber \\
 & &\hspace{-.5cm} \times \left[ \text{Erf}\left( \frac{i}{2\sqrt{2 \tilde
\ell}} (\tilde f + 4 \tilde \ell + 2 \omega)\right)-\text{Erf}\left(
\frac{i}{2\sqrt{2 \tilde \ell}} (\tilde f - 4 \tilde \ell + 2 \omega)
\right)\right] \, .
\end{eqnarray}

\bibliographystyle{bj}
\bibliography{stretch}

%\end{document}

\newpage


\end{document}